\documentclass[prl,superscriptaddress,twocolumn,showpacs]{revtex4}

\usepackage{graphicx}

\begin{document}

\title{Coherent Eavesdropping Attacks in Quantum Cryptography:\\
Nonequivalence of Quantum and Classical Key Distillation}

\author{Dagomir Kaszlikowski}
\affiliation{Department of Physics, National University of Singapore,
Singapore 117\,542, Singapore}

\author{Jenn Yang Lim}
\affiliation{Department of Physics, National University of Singapore,
Singapore 117\,542, Singapore}

\author{Leong Chuang Kwek}
\affiliation{National Institute of Education, 
Nanyang Technological University, Singapore 639\,798, Singapore}
\affiliation{Department of Physics, National University of Singapore,
Singapore 117\,542, Singapore}

\author{Berthold-Georg Englert}
\affiliation{Department of Physics, National University of Singapore,
Singapore 117\,542, Singapore}

\date{20 December 2003}

\begin{abstract}
The security of a cryptographic key that is generated by communication 
through a noisy quantum channel relies on the ability to distill a
shorter secure key sequence from a longer insecure one. 
We show that 
--- for protocols that use quantum channels of any dimension and completely
characterize them by state tomography --- 
the noise threshold for classical advantage distillation is substantially
lower than the threshold for quantum entanglement distillation because the
eavesdropper can perform powerful coherent attacks. 
The earlier claims that the two noise thresholds are identical, which were
based on analyzing incoherent attacks only, are therefore invalid.
\end{abstract}


\maketitle

%%%%%%%%%%%Introductory remarks%%%%%%%%%%%%%%%%%%%%%%%

Unavoidably, all practical implementations of protocols for quantum
cryptography suffer from noise in the quantum channel and must, therefore,
face the problem of generating a secure cryptographic key from noisy raw data.
Since the noise can wholly result from eavesdropping attacks, the central
question is: 
Where is the noise threshold below which the communicating parties, 
Alice and Bob, can generate a secure key at all?

Except at rather low noise levels, the extraction of a shorter secure key
sequence from a longer, noisy and insecure, raw key sequence must involve a
\emph{distillation} procedure, either a variant of ``quantum entanglement
distillation'' (QED \cite{QED}) or of ``classical advantage distillation''
(CAD \cite{CAD}). 
A recent paper asserts that, for an important class of protocols,   
``the thresholds for QED and CAD are the same'' so that 
``the two distillation procedures are equivalent in the
sense that neither offers a security advantage over the other'' 
\cite{ADvsED}, and the same conclusion has also been reached 
in Ref.~\cite{AGS03}, and earlier in Ref.~\cite{GW99} for qubit protocols.

But the analysis in Refs.~\cite{ADvsED,AGS03,GW99} takes only
\emph{incoherent} attacks into account, in which eavesdropper
Eve acquires as much information as possible about each individual entry in
the raw key data shared by Alice and Bob. 
This raw key is, however, of secondary interest to Eve, who primarily wishes 
to know well the final, distilled key.
Accordingly, she might be better off by satisfying her
primary interest directly, rather than first maximizing her transitory
knowledge of secondary interest. 

Indeed, we report here that Eve can gain more knowledge about the final key
sequence by a suitable \emph{coherent} attack, if Alice and Bob make use of
CAD. 
The actual noise threshold for CAD is therefore lower than the one found in
Refs.~\cite{ADvsED,AGS03,GW99}. 
And since the threshold for QED is independent of Eve's eavesdropping
strategy, it follows that the two distillation procedures are not equivalent: 
The noise level in a quantum channel can be above the CAD threshold but below
the QED threshold. 

In this context, then, coherent eavesdropping attacks truly outperform 
incoherent attacks. 
As plausible as this may seem in hindsight, it is not at all obvious.
In fact, when neither QED nor CAD are performed, it has been argued that
coherent attacks cannot be more powerful than incoherent attacks \cite{GC+W}.

%%%%%%%%%%%%Description of tomographic protocol%%%%%%%%%%%%%%%%%%%%%%%%%%%%%%5

We reconsider the tomographic protocol for quantum cryptography of
Refs.~\cite{TomoCrypt} and \cite{ADvsED}.
A source distributes pairs of \emph{qunits} ($n$-dimensional quantum objects,
$n\geq2$) to Alice and Bob, and they measure nondegenerate observables that
are randomly chosen from a tomographically complete set of $n+1$ observables. 
Both keep a record of the observable they have
measured for each pair and of the measurement result. 
As in Refs.~\cite{TomoCrypt,ADvsED} we denote by $|m_k\rangle$ the $k$th
eigenket of Alice's $m$th observable and by $|\overline{m}_k\rangle$ the $k$th
eigenket of Bob's $m$th observable.  
The correspondence between the orthonormal bases associated with the
observables is established by requiring that
\begin{equation}
\braket{0_j}{m_k} =\braket{\overline{m}_k}{\overline{0}_j}
\label{eq:1}
\end{equation}
holds for $j,k=0,1,2,\dots,n-1$ and $m=0,1,2,\dots,n$.
We are thus pairing Alice's $m$th observable with Bob's $m$th observable.

This pairing is essential in defining the two-qunit state
\begin{equation}\label{eq:2}
|\psi\rangle = \frac{1}{\sqrt n}\sum_{k=0}^{n-1}\ket{m_k\overline{m}_k}
\end{equation}
that Alice and Bob wish to receive from an ideal source.
The two-qunit kets $\ket{m_k\overline{m}_k}$ refer to the $m$th pair of
observables but, as a consequence of \Eqref{1}, $\ket{\psi}$ is the same
regardless of the $m$ value chosen in \eqref{2}.

When the transmission is over, Alice and Bob publicly announce their choice of
observables, their respective $m$ values, for all qunits, while keeping the
measurement results, their \emph{nit values}, secret. 
The qunit pairs constitute two groups, one in which the measurement bases
match (both $m$ values are the same, which happens with probability
$1/(n+1)$), and the other in which the bases do not match.  
In the absence of noise, the nit values of the first group are
perfectly correlated and thus give rise to a cryptographic key in an alphabet
with $n$ letters. 

In reality, however, Alice and Bob must take into account Eve's attempts at
eavesdropping and the resulting disturbance of the quantum channel. 
As a consequence, the source effectively emits qunit pairs whose properties
are described by a statistical operator $\rho$ that differs from the ideal
projector $\ketbra{\psi}$.
Since Alice and Bob measure tomographically complete sets of observables on
their respective qunits, they can determine the actual $\rho$ from their
measurement results.  
They exploit all data of the mismatched bases for this purpose, and some of
the matched-bases data.  
Ideally, they wish for $\rho=|\psi\rangle\langle\psi|$ but, realistically,
they expect to find a $\rho$ of the form 
\begin{equation}
\rho = \ket{\psi}(\beta_0-\beta_1)\bra{\psi} + \frac{\beta_1}{n}
 = \ket{\psi}(1-\mathcal{E})\bra{\psi} + \frac{\mathcal{E}}{n^2}\,,
\label{eq:3}
\end{equation}
with $\beta_0+(n-1)\beta_1=1$ and $\mathcal{E}=1-(\beta_0-\beta_1)=n\beta_1$. 
This is what one gets when an imperfect transmission line admixes unbiased
noise to $\ketbra{\psi}$, the fraction of the admixture being
quantified by the noise parameter $\mathcal{E}$ \cite{Fidel}.  
The nonnegative parameters $\beta_0$ and $\beta_1$ have the following physical
significance: $\beta_0$ is the probability that Alice and Bob get the same nit
value when the bases match, and $\beta_1$ is the probability that Bob gets a
particular one of the $n-1$ values that are different from Alice's nit value.

The relevant range of parameters is such that
$0\leq\beta_1\leq\beta_0\leq1$ or $0\leq \mathcal{E}\leq1$ 
as only then the state $\rho$
has the interpretation of an admixture of noise to the pure state
$\ket{\psi}$. 
The limiting values mark the extreme situations of
``no noise at all'' ($\beta_0=1$, $\beta_1=0$, $\mathcal{E}=0$) and 
``nothing but noise'' ($\beta_0=\beta_1=1/n$, $\mathcal{E}=1$).

Sources that emit two-qunit states $\rho$ of a kind different from the one in
\Eqref{3} are not regarded as trustworthy by Alice and Bob. 
As the crucial, defining step of the \emph{tomographic} protocol, 
they accept the source only if their state tomography confirms that the
source emits statistically independent qunit pairs with a $\rho$ of
the form \eqref{3}.  
Otherwise, they switch to another source \cite{KP}.

%%%%%%%%%%%%EAVESDROPPING%%%%%%%%%%%%%%%%%%%%%

We grant Eve full control over the two-qunit source.
Then, in order to acquire as much information about the
key as possible, she entangles one ancilla each with the qunit pairs
sent to Alice and Bob. 
Since they must receive the pairs in the state \eqref{3}, Eve's choices are
severely limited.
She is bound to prepare entangled two-qunit--ancilla states 
with a ket of the form
\cite{TomoCrypt} 
\begin{eqnarray}
\ket{\Psi}&=&
\sqrt{\frac{\beta_0}{n}}\sum_{k=0}^{n-1}\ket{m_k\overline{m}_k}
\ket{E_{kk}^{(m)}}  \nonumber\\&&
+\sqrt{\frac{\beta_1}{n}}\sum_{k\neq l}\ket{m_k\overline{m}_l}
\ket{E_{kl}^{(m)}}\,, \label{eq:4}
\end{eqnarray}
where the $\ket{E_{kl}^{(m)}}$ are normalized ancilla kets. 
The sets of ancilla states pertaining to different values of $m$ 
are unitarily equivalent. 
For a given $m$ value, the ancilla states $\ket{E_{kl}^{(m)}}$ with 
$k\neq l$ are orthogonal to each other and orthogonal to the ones with $k=l$.
The latter are not orthogonal among themselves (except when
$\beta_0=\beta_1$, the case of pure noise and of very little
interest), but rather have the same inner products for all pairs, i.e.,
\begin{equation}
\braket{E^{(m)}_{kk}}{E^{(m)}_{ll}}
=1-\frac{\beta_1}{\beta_0}
=\frac{1-\mathcal{E}}{1-(1-1/n)\mathcal{E}}
 \equiv \lambda \label{eq:5}
\end{equation}
for $k\neq l$.

%%%%%%%%%mutual information%%%%%%%%

Alice and Bob can generate a secure key if the correlations between their nit
values (i.e.\ their measurement results for matched bases) are stronger than
the correlations between, say, Alice's values and the values that Eve obtains
by whatever measurements on the respective ancillas. 
In technical terms, the mutual information between Alice and Bob must be
larger than the mutual information between Alice and Eve \cite{CK}.
The efficiency of the protocol is proportional to the difference between the
two mutual information values.

If $\mathcal{E}$ is sufficiently small,
this condition is already met for the raw key
sequence (see Ref.~\cite{Pyra} for the actual criterion),  
and then the generation of the secure key is a matter of applied
coding theory. 
Alice and Bob know if this is the case because they 
have determined the actual value of $\mathcal{E}$ 
by the two-qunit--state tomography.

If $\mathcal{E}$ is found to be not ``sufficiently small,'' 
then the raw data is too noisy and Alice and Bob
must use a distillation procedure to improve the situation, 
either QED or CAD; 
see Refs.~\cite{QED} and \cite{CAD}, respectively.  
When employing QED, Alice and Bob process the qunit pairs before measuring
the observables that give the nit values and so form, in essence, 
a purified set of qunit pairs with a new $\mathcal{E}$ value that 
is small enough.
As established in Ref.~\cite{QED}, QED can be performed successfully if
$\beta_0>2\beta_1$ and only then.
In terms of the noise parameter $\mathcal{E}$ this means
\begin{equation}
  \label{eq:6}
  \mathcal{E}<\mathcal{E}^\mathrm{(QED)}_\mathrm{th}=\frac{n}{n+1}\,,
\end{equation}
which thus identifies the noise threshold for QED.
The two-qunit state of \eqref{3} is separable if 
$\mathcal{E}\geq \mathcal{E}^\mathrm{(QED)}_\mathrm{th}$, 
so that QED exhausts the full range of
$\mathcal{E}$ values for which cryptographic security is potentially possible,
and no other procedure can ever have a larger $\mathcal{E}$ range.

Whereas the qunits are manipulated in QED, one processes the measured nit
values when performing CAD.
Therefore, the implementation of QED is a very challenging hardware problem
whereas rather simple software is needed for CAD. 
This practical advantage of CAD over QED comes, however, at a price.
As we now proceed to demonstrate, the noise threshold for CAD is lower than
the QED threshold of \Eqref{6}.   

In the CAD protocol Alice and Bob divide their raw key sequence of nit values
(for the matched bases) into blocks of length $L$. 
For each block Alice tosses an $n$-sided die and adds, modulo $n$, the
resulting random value to each value of the block. 
She so obtains a new block, which she sends to Bob through an
authenticated but insecure public channel. 
After receiving the block, he subtracts his corresponding block from it 
(modulo $n$).
If all the nit values are the same after the subtraction, which happens with
probability $\beta_0^L+(n-1)\beta_1^L$, Bob informs Alice that this is a good 
block, otherwise it is a bad block \cite{KP}.

All good blocks together define a distilled sequence of nit values, one value
for each block:
Alice records the random nit values she added, Bob the nit values he found
after the subtraction.
The distilled sequence can be characterized by probabilities
$\beta_0^{(L)}$ (Bob has the same value as Alice) and $\beta_1^{(L)}$
(he has a particular one of the $n-1$ other values)  
that are the $L>1$ analogs of their $L=1$ versions in \Eqref{3} 
and related to them by~\cite{ADvsED}
\begin{equation}
  \label{eq:7}
  \frac{\beta_1^{(L)}}{\beta_0^{(L)}}
  =\left(\frac{\beta_1}{\beta_0}\right)^L\,.
\end{equation}
Accordingly, with growing $L$, the distilled sequence has exponentially less
noise than the raw sequence. 

After performing CAD, the resulting mutual information between Alice and Bob 
is given by
\begin{equation}
I_L(A\&B) = 1+\beta_0^{(L)}\log_n\beta_0^{(L)}
+(1-\beta_0^{(L)})\log_n\beta_1^{(L)} 
\label{eq:8}
\end{equation}
with  
\begin{equation}
\bigl(\beta_0^{(L)},\beta_1^{(L)}\bigr) =
\frac{\bigl(\beta_0^L,\beta_1^L\bigr)}{\beta_0^L+(n-1)\beta_1^L}\,,
\label{eq:9}
\end{equation}
where, for convenience, we measure information in nits ($\log_n$) rather 
than bits ($\log_2$).
The asymptotic forms
\begin{eqnarray}
  \label{eq:10}
&& \beta_0^{(L)}\simeq1-(n-1)\bigl(\beta_1/\beta_0\bigr)^L\,,\quad
  \beta_1^{(L)}\simeq\bigl(\beta_1/\beta_0\bigr)^L\,,\nonumber\\
&&I_L(A\&B)\simeq1-(n-1)\bigl(\beta_1/\beta_0\bigr)^L\log_n
           \bigl(\beta_0/\beta_1\bigr)^L
\end{eqnarray}
apply for $L\gg1$, so that the difference $1-I_L(A\&B)$ decreases
exponentially with increasing block length $L$.

Eve's strategy is as follows. 
She stores her ancillas and waits passively until Bob announces his approval
or rejection of the given block to Alice over the public channel. 
The ancillas of the bad blocks are then of no further interest.
For each good block Eve knows that either (I)
all corresponding nit values in Alice's and Bob's blocks are the same, or (II)
they differ by the same amount (modulo $n$).
There is no room for any other possibility in the CAD protocol.
For instance, Alice could have the block $0121$ for $L=4$ and $n=3$,  
and then there are three possible blocks for Bob, namely $0121$, $1202$,
or $2010$,
resulting from Alice's addition of $0$, $1$, or $2$, respectively. 
The fraction $\beta_0^{(L)}$ of the good blocks are case-I blocks, the
fraction  $(n-1)\beta_1^{(L)}=1-\beta_0^{(L)}$ are case-II blocks.  

For each good block, Eve has a corresponding set of ancilla states. 
Rather than measuring the ancillas one-by-one (incoherent attack), she
performs a joint measurement on all $L$ of them (coherent attack) to acquire
knowledge about the value that, say, Alice assigns to the block.
In case (II), Eve knows exactly Alice's and Bob's nit values
because all the respective ancilla states $\ket{E_{kl}^{(m)}}$ 
have $k\neq l$ and are thus orthogonal to all other potential ones 
(recall the remark after \Eqref{4}) and can be distinguished unambiguously. 

In case (I), Alice and  Bob have an identical block of nit values to begin
with, and Eve's ancillas states are all of the $k=l$ kind.
Although she can establish easily that the blocks are of case (I), 
she cannot distinguish the potential ancilla states unambiguously because they
are not orthogonal to each other.
Accordingly, Eve has no certain knowledge of the distilled nit values for
case-I blocks.
But by making good use of the classical information that is exchanged
publicly between Alice and Bob during the distillation process, Eve can learn
a lot about these nit values.
In fact, she only needs to distinguish $n$ possible $L$-ancillas
states, and they are almost orthogonal to each other when $L\gg1$.

The situation is best illustrated with an example.
Suppose and Alice and Bob have the same block $0121$ for $n=3$ and $L=4$,  
and her random nit value is $1$.
After addition (modulo $3$), she sends the processed block $1202$ to Bob
via a public channel. 
Eve, who is fully knowledgeable of all such broadcast information and has
already established that she is dealing with a case-I block, then infers
that the unprocessed block is either $1202$, or $0121$, or $2010$, and the
distilled values would be $0$, $1$, and $2$, respectively. 
She concludes that the four ancillas in question are in the
four-ancilla state with ket
$\ket{E_{11}E_{22}E_{00}E_{22}}$, or $\ket{E_{00}E_{11}E_{22}E_{11}}$,
or $\ket{E_{22}E_{00}E_{11}E_{00}}$, where the $E_{kk}^{(m)}$'s at the same
positions have identical $m$ values that we leave implicit. 
Any two of these four-ancilla states have the same inner product of
$\lambda^4$, inasmuch as 
\begin{eqnarray*}
&&\braket{E_{11}E_{22}E_{00}E_{22}}{E_{00}E_{11}E_{22}E_{11}}
\nonumber\\
&=&\braket{E_{11}}{E_{00}}\braket{E_{22}}{E_{11}}\braket{E_{00}}{E_{22}}
\braket{E_{22}}{E_{11}}=\lambda^4\,,  
\end{eqnarray*}
for instance.
More generally, for each case-I block of length $L$, Eve needs
to distinguish  $n$ possible $L$-ancilla states, with inner products 
of $\lambda^L$ for each pair of states.

For large $L$, this inner product is very small and, therefore \cite{Pyra},
Eve maximizes her mutual information with Alice (or Bob) by the so-called 
``square-root measurement'', which is always the error-minimizing measurement.
Her probability of inferring a distilled case-I nit value correctly is then
given by~\cite{TomoCrypt}
\begin{equation}\label{eq:11}
\eta_0^{(L)} =
\left(\frac{\sqrt{1+(n-1)\lambda^L}+(n-1)\sqrt{1-\lambda^L}}{n}\right)^2\,,
\end{equation}
and she gets a particular one of the $n-1$ wrong values with probability
\begin{equation}
  \label{eq:11a}
  \eta_1^{(L)}=\frac{1-\eta_0^{(L)}}{n-1}
=\left(\frac{\sqrt{1+(n-1)\lambda^L}-\sqrt{1-\lambda^L}}{n}\right)^2\,.
\end{equation}
The resulting mutual information between Alice and Eve is
\begin{eqnarray}
I_L(A\&E) &=& 1 + \beta_0^{(L)}\left( \eta_0^{(L)}
\log_n{\eta_0^{(L)}}\right.  \nonumber\\
&& \left. {}+(1-\eta_0^{(L)})\log_n\eta_1^{(L)}\right)\,.
\label{eq:12}
\end{eqnarray}
The asymptotic forms
\begin{eqnarray}
  \label{eq:13}
&& \eta_0^{(L)}\simeq1-\frac{1}{4}(n-1)\lambda^{2L}\,,\quad
  \eta_1^{(L)}\simeq\frac{1}{4}\lambda^{2L}\,,\nonumber\\
&&I_L(A\&E)\simeq1-\frac{1}{4}(n-1)\lambda^{2L}\log_n
           \bigl(1/\lambda\bigr)^{2L}
\end{eqnarray}
apply for $L\gg1$, so that the difference $1-I_L(A\&E)$ also decreases
exponentially with increasing block length $L$.

%%%%%%%CK SECURITY THRESHOLD%%%%%%%%%%%

Now, according to the Csisz\'ar-K\"orner Theorem \cite{CK}, 
a secure cryptographic key can be generated from the
raw key sequence, by means of a suitably chosen error-correcting
code and classical (one-way) communication between Alice and Bob,
if the mutual information between Alice and Bob exceeds that
between Eve and either of them. 
This ensures success of the CAD procedure whenever $I_L(A\&B)>I_L(A\&E)$
obtains. 

Upon comparing the large-$L$ versions of the mutual informations in \eqref{10}
and \eqref{13}, we note that, for sufficiently long blocks, successful CAD is
assuredly possible if $\beta_1/\beta_0<\lambda^2$.  
Since $\lambda+\beta_1/\beta_0=1$,
the corresponding criterion for the noise level is 
\begin{equation}
  \label{eq:15}
 \mathcal{E}<\mathcal{E}^\mathrm{(CAD)}_\mathrm{th}
=\frac{n}{n+(1+\sqrt{5})/2}\,.
\end{equation}
The CAD threshold value thus identified is always lower than the 
QED threshold value of \Eqref{6}, because the golden mean $(1+\sqrt{5})/2$
exceeds unity.

\begin{figure}[!t]
\centering\rule{0pt}{4pt}\par
\includegraphics{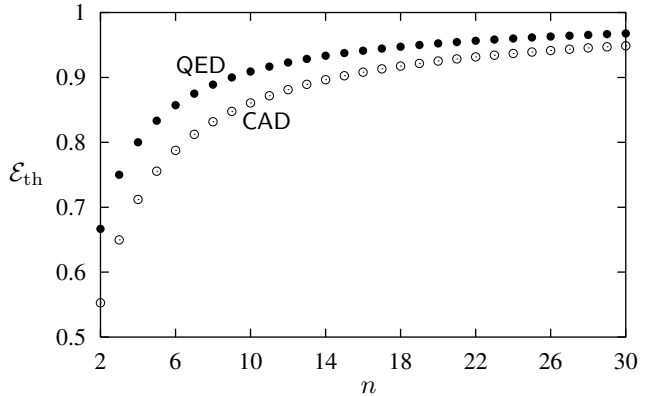}
\caption{\label{fig:thresholds} %
Noise thresholds $\mathcal{E}_\mathrm{th}$ for 
Quantum Entanglement Distillation (QED, full circles) 
and Classical Advantage Distillation (CAD, empty circles) under
coherent eavesdropping attacks, as a function of the dimension $n$ of the
transmitted quantum objects. 
QED can tolerate substantially more noise in the channel. }
\end{figure}

Figure \ref{fig:thresholds} shows the two noise thresholds as a function of
$n$, for $2\leq n \leq30$. 
It can be seen clearly that QED can tolerate substantially more noise 
in the channel than CAD, in particular in the qubit case of $n=2$,
where the thresholds are at ${\mathcal{E}=2/3=66.7\%}$ and 
${\mathcal{E}=1-\sqrt{1/5}=55.3\%}$,
respectively.

In summary, we have established that Eve has a real advantage from coherent
attacks if Alice and Bob perform CAD.
The coherent attack that we describe in detail aims at getting optimal
knowledge about each nit value of the distilled key individually.
It is conceivable (but we do not consider it likely) that more involved
coherent attacks, which would provide knowledge about groups of distilled-key
nit values, are even more powerful. 
Strictly speaking, the threshold stated in \Eqref{15} must, therefore, be
regarded as an upper bound on the noise threshold for CAD.

We note further that other procedures for advantage distillation are also
vulnerable to coherent attacks of the considered kind.
This is true, in particular, for the parity-check distillation for qubit
protocols, for which an analogous coherent attack can be
analyzed easily \cite{parity}.

We are indebted to Artur Ekert, Ajay Gopinathan, Christian Kurtsiefer, 
and Yeong Cherng Liang for sharing their insights with us.
This work was supported by A$^*$Star Grant No.\ 012-104-0040.


\begin{thebibliography}{99}
\bibitem{Fidel}
The fidelity of the corresponding effective quantum channel between Alice and
Bob is $\mathcal{F}=1-(1-1/n^2)\mathcal{E}$.
\bibitem{KP}
Eve is, of course, aware of this because, 
as required by the Kerkhoff Principle of cryptology, she knows all procedural
rules of the protocol executed by Alice and Bob and has access to all
information they exchange publicly.
\end{thebibliography}
\end{document}